%% file: p_17_icra_switched_hyq.tex
\documentclass[letterpaper, 10 pt, conference]{ieeeconf}  

\IEEEoverridecommandlockouts                              

\usepackage{graphics} 
\usepackage{epsfig} 
\usepackage{mathptmx} 
\usepackage{times} 
\usepackage{amsmath} 
\usepackage{amssymb}  
\usepackage{todonotes}
\usepackage{float}
\usepackage{url}
\usepackage[noadjust]{cite}
 
\input{mathdef}

\title{\LARGE \bf
An Efficient Optimal Planning and Control Framework For Quadrupedal Locomotion 
}

\author{Farbod Farshidian$^*$, Michael Neunert$^*$, Alexander W. Winkler$^*$, Gonzalo Rey$^{\dagger}$, Jonas Buchli$^*$
\thanks{$^*$All authors are with the Agile \& Dexterous Robotics Lab, ETH Z\"urich, Switzerland, email: \{farbodf, neunertm, winklera, buchlij\}@ethz.ch}%
\thanks{$^{\dagger}$Gonzalo Rey is with Moog grey@moog.com}%
}

\begin{document}

\maketitle
\thispagestyle{empty}
\pagestyle{empty}

\begin{abstract}
In this paper, we present an efficient Dynamic Programing framework for optimal planning and control of legged robots. First we formulate this problem as an optimal control problem for switched systems. Then we propose a multi--level optimization approach to find the optimal switching times and the optimal continuous control inputs. Through this scheme, the decomposed optimization can potentially be done more efficiently than the combined approach. Finally, we present a continuous-time constrained LQR algorithm which simultaneously optimizes the feedforward and feedback controller with $O(n)$ time-complexity. In order to validate our approach, we show the performance of our framework on a quadrupedal robot. We choose the Center of Mass dynamics and the full kinematic formulation as the switched system model where the switching times as well as the contact forces and the joint velocities are optimized for different locomotion tasks such as gap crossing, walking and trotting. 
\end{abstract}

\section{INTRODUCTION}
Motion control in robotics promises to bring more autonomy to robots in the sense that they can plan and control their motion in the environment without or with minimum operator interference. In our terminology, motion control refers to the whole process of planning and control of the motion of a robot over the course of time. Due to the hybrid and nonlinear nature of the legged locomotion problem, finding an optimal plan and controller is a challenging task. Furthermore, a practical implementation of a motion control framework requires to deal with uncertainty and dynamic changes of the environment. The Model Predictive Control (MPC) approach can address these issues to some extent. MPC is an optimal control framework which repeatedly plans the system's motion in the future and then only executes the first part of the plan until new information becomes available.

Among the different ways to solve the MPC problem, using optimal feedback planners is one of the most promising approaches. In contrast to conventional methods where only the feedforward plan is calculated, the feedback planners design the stabilizing controller parallel to the open--loop plan. Using feedback planners is more essential on the high dimensional problems where the MPC loop runs slower than the controller rate. These feedback planners are mostly based on the Dynamic Programming (DP) framework rather than the widely used Trajectory Optimization (TO) approaches. However, known as the curse of dimensionality, DP does not naturally scale to high dimensional systems. Therefore, for a long time, it has been the belief that a DP-based algorithm cannot solve real-life, high dimensional problems such as legged robot locomotion. However, the recent progress in efficient DP-based algorithms as well as the advent of fast and cheap processors have brought attention back to the DP approaches once again. While algorithms like Mayne's DDP (Differential Dynamic Programming) have existed since 1966 \cite{mayne66}, it was only until recently, that the fast DP-based algorithms were revisited \cite{todorov05,sideris05} and their performance were demonstrated in different robotic platforms such as quadrotors, swimming robots, and legged robots. 

This class of fast and efficient algorithm are formally known as Sequential, Linear, Quadratic (SLQ) methods. In order to avoid the curse of dimensionality which arises from the value function calculation, it uses a local quadratic approximation of the value function to calculate its value in the vicinity of the current operating points. Then, each iteration of this approximation is followed by a forward integration of the system dynamics in order to update the nominal operating points for the next iteration. In spite of their efficiency, the SLQ type algorithms have a major drawback in that they often cannot efficiently handle equality or inequality constraints. In this contribution, we address this issue by proposing a method to formulate legged robot locomotion as a constrained switched system optimal control problem which can be solved via our efficient feedback planner.


\subsection{Literature Review}
Broadly speaking, there are two basic approaches for solving the optimal motion control problem namely Dynamic Programming and Trajectory Optimization. The DP approach is based on the principle of optimality. The methods in this framework normally break the optimal control problem into a collection of the simpler subproblems, solve each of those subproblems just once, and store their solutions. 
On the other hand, the TO methods are techniques for computing open--loop solutions to the optimal control problem. TO transforms the original infinite--dimension continuous problem into a finite dimensional Nonlinear Programing (NLP). 

TO can be categorized into direct and indirect methods \cite{von92}. The indirect methods which are based on the Pontryagin's maximum principle have a limited application in robotics. Direct methods can be divided into three main groups namely direct single shooting, direct multiple shooting \cite{diehl06} and direct transcription \cite{enright92}. The single shooting approach starts by discretizing the control inputs, then performs a forward integration through an ODE solver. Due to the unappealing numerical characteristics of this method, its application on problems with long optimization horizon is rather limited.    

The multiple shooting approach is based on a similar idea as single shooting with the difference that it divides the long integration into smaller pieces. This technique helps to reduce the cumulative affect of the early parameters of trajectory on the later ones which results in a better NLP problem. However, in order to ensure continuity, it adds a set of matching conditions at each interval. This method has been employed successfully for trajectory planning on a number of locomotion tasks \cite{koch12, lengagne13}. On the other hand, the direct transcription methods use an approximation scheme to transform the optimal control problem to a finite dimension NLP. To do so, the direct transcription methods discretize the state and control trajectories into a finite number of nodes and then interpolate in between the nodes by spline approximation. In \cite{posa14, pardo15} a direct transcription approach is applied to legged robot motion control where the end-effector placement is incorporated into the whole-body planning. 

Regardless of the chosen method, many of these approaches have shown their capabilities in planning contact rich motion on legged robots. However, the transformation of the optimal control problem to a general NLP introduces a high dimensional optimization problem. In addition to slow convergence rates and a complex objective landscape, the outcome of TO trajectories cannot be implemented directly on hardware and in most cases require a tracking controllers to implement the designed trajectories \cite{ihmc13,winkler17}. This can produce non-optimal motion in real hardware experiments.   


Recently, there has been an increasing interest in efficient DP-based feedback planning approaches where the feedforward plan is designed together with the feedback controller \cite{mayne66,sideris05,todorov05}. DP--based approaches have been applied for controlling a humanoid \cite{tassa12} and a quadruped robot \cite{neunert16}. Both applications use smooth contact models in order to discover the contact sequence as well as the motion trajectories and stabilizing feedback controller. This smooth contact assumption is required since none of such methods is able to deal with the state--input constraints introduced by contacts. There are few examples of constrained DP--based methods that can potentially scale to the legged robotics problem \cite{ohno78, yakowitz86,romano15}. However, they either need a near--optimal initial guess to converge \cite{ohno78,yakowitz86}, or are computationally inefficient due to an extra backward pass in each iteration of algorithm \cite{romano15}.  

\subsection{Contributions}
In this contribution, we introduce an approach to transform the optimal motion planning and control problem for legged robots into a more tractable optimization problem based on a multi--level optimization algorithm. To do so, we reformulate this problem into an optimal control problem for switched systems assuming that the switching mode sequence is given. In the new problem, the optimization variables are the original system's control inputs as well as the switching times between different modes of motion. Based on this model, we use a multi--level optimization approach introduced in \cite{farshidian16} which optimizes the switching time and the continuous control inputs in two alternating optimization problems. Furthermore, we introduce a continuous-time, constrained SLQ algorithm with state and input equality constraints which has a $O(n)$ time--complexity. The idea of constrained SLQ has originally been introduced in~\cite{sideris10} for the discrete--time formulation. In our work, we extend this work to the continuous case while we reduce its complexity to $O(n)$. Furthermore, the continuous time version allows us to use adaptive step-size integrators, which helps to achieve shorter runtimes in practice. Finally, we evaluate our framework performance on a quadruped robot where we use the centroidal dynamics and full kinematics model for motion planning and control.    

\section{Problem formulation}

\subsection{Switched System Formulation}
In this section, we investigate an approach to optimize motion control for a legged robot. While our main focus is on legged robots, the discussion can be extended to any hybrid system with finite switching modes such as a manipulator which makes contact with environment. In each mode $i$, the equation of motion of a legged robot can be described as a second order system with a set of equality constraints introduced by the contacts. 
\vspace{-1em}

\begin{small} 
\begin{align} 
&\vM(\vq) \ddot{\vq} + \vh(\vq,\dot{\vq}) = \vtau + \sum\nolimits_{c\in\mathbb{C}_i}{\vJ_c^T \vlambda_c } \notag \\
& \begin{cases}
    \dot{\vd}_c^i = 0       & \, c\in\mathbb{C}_i \\
    \vlambda_c=\mathbf{0}   & \, c\in\overline{\mathbb{C}}_i
\end{cases}  
\qquad (\vq,\dot{\vq}) \notin S_{ij}
\label{eq:hybrid_predefined}
\end{align} 
\end{small}%
where $\vq$ is the generalized coordinate consisting of the base pose and joint angles, $\tau$ is the generalized torque vector, and $\vlambda_c$ is the force vector at contact point $c$. $\vM$, $\vh$, and $\vJ_c$ are respectively Inertia matrix, Coriolis--gravity forces, and Jacobian associate to contact point $c$. $\vd_c^i$ is the relative distance of the contact point $c$ to the current contact surfaces. The set of active contact forces in mode $i$ is $\mathbb{C}_i$ and $S_{ij}$ is a switching surface which transforms the system dynamics form a current mode $i$ to the next mode $j$. Normally, such transition (e.g. foot touch--down) causes a discontinuity in the state of the robot. However, to simplify the problem we will assume that this discontinuity is negligible and the resulting impact force will be dealt with by a suitable feedback controller. Furthermore, we assume that the gait sequence is predefined. This information can be provided either by restricting the motion to a specific gait or through a discrete search over the possible gait sequences. 

Even with a given gait sequence, optimizing over the hybrid system in Equation~\eqref{eq:hybrid_predefined} is a challenging problem. The reason is that it requires to optimize indirectly over the mode switches by affecting the state trajectory through input controls. However, this problem can be transformed to a simpler switched system optimization problem by introducing extra control variables, namely the switching times $\{s_i\}_{i=0}^I$ and adding some additional constraints. Here $I$ is the total number of the modes. Therefore we get
\vspace{-1em}

\begin{small}
\begin{align} 
&\vM(\vq) \ddot{\vq} + \vh(\vq,\dot{\vq}) = \vtau + \sum\nolimits_{c\in\mathbb{C}_i}{\vJ_c^T \vlambda_c } \notag \\
& \begin{cases}
    \dot{\vd}_c^i = 0        & \, c\in\mathbb{C}_i \\
    \vlambda_c=\mathbf{0}  & \, c\in\overline{\mathbb{C}}_i  \\
	\vd_c^j > 0        & \, c\in\overline{\mathbb{C}}_i 
\end{cases}  
\qquad t \in [s_{i-1},s_i)
\label{eq:switched_equations}
\end{align} 
\end{small}%
where $\vd_c^j$ is the distance to the switching surface of the next mode $j$. Through this technique, we can transform the hybrid system with predefined mode switches to a switched system with additional inequality constraints. A major advantage of this approach is that we can use the well-developed and efficient algorithms from the switched system optimal control literature instead of its hybrid systems counterpart. In the next section, we discuss how to formulate and solve the optimal control problem for such a switched system.  

\subsection{Optimal Control for Switched System}
The optimal control problem for a system defined in Equation~\eqref{eq:switched_equations} can be formulated as follows
\vspace{-1em}

\begin{small}
\begin{equation}
\label{eq:general_op}
\begin{aligned}
& \underset{\{s_i\}, \vu (\cdot)}{\text{minimize}}
& & \sum\limits_{i=0}^{I-1} {\Phi_i(\vx (s_{i+1})) + \int_{s_i}^{s_{i+1}} L_i(\vx, \vu) dt }  \\
& \text{subject to}
& & \dot{\vx} =  \vf_i(\vx, \vu), \hspace{4mm} \vx(s_0) = \vx_0, \quad \vx(s_i^-) = \vx(s_i^+) \\
& & & \vg_1(\vx, \vu, t) = 0, \  \vg_2(\vx, t) = 0,
\end{aligned}
\end{equation}
\end{small}%
where $\{s_i\}$ and $\vu (\cdot)$ are respectively the switching times and the continuous--time control input vector. For each mode $i$, the nonlinear cost function consists of a terminal cost and an intermediate cost. $\vf_i(\cdot)$ is the system dynamics in mode $i$. $\vg_1(\cdot)$ and $\vg_2(\cdot)$ are the state-input and pure state constraints. The optimization defined in Equation~\eqref{eq:general_op} is an optimal control problem for a switched system. 

In this paper, we use an algorithm which is based on the idea of multi-level optimization first introduced by Xu and Antsaklis \cite{xu04}. In this approach, the procedure of synthesizing an optimal control law for a switched system is divided into two subtasks: 1) finding the optimal switching times between consecutive modes, 2) optimizing the continuous controls. Based on this, we reformulate \eqref{eq:general_op} as 
\vspace{-1em}

\begin{small}
\begin{equation*}
\begin{aligned}
& \underset{\{s_i\}}{\text{minimize}}
& & \sum\limits_{i=0}^{I-1} {\Phi_i(\vx^* (s_{i+1})) + \int_{s_i}^{s_{i+1}} L_i(\vx^*, \vu^*) dt }  \\
& \text{subject to}
& & \dot{\vx}^* =  \vf_i(\vx^*, \vu^*), \ \vx(s_0) = \vx_0, \ \vx^*(s_i^-) = \vx^*(s_i^+) \\
& & & \vu^*(\cdot) = \argmin \Big\{ \sum\limits_{i=0}^{I-1} {\Phi_i(\vx (s_{i+1})) + \int_{s_i}^{s_{i+1}} L_i(\vx, \vu) dt } \Big\} \\
& & &
	\begin{aligned}
	\text{subject to \quad} & \dot{\vx} =  \vf_i(\vx, \vu), \ \vx(s_0) = \vx_0, \ \vx(s_i^-) = \vx(s_i^+) \\
	& \vg_1(\vx, \vu, t) = 0, \ \vg_2(\vx, t) = 0. 
	\end{aligned}
\end{aligned}
\end{equation*}
\end{small}%
In \cite{xu04}, this optimization is reformulated by augmenting the state vector with an extra state which represents the switching times. Then it uses two alternating optimization procedures to find the optimal switching times and the continuous--time controllers. By fixing the switching time state (in the bottom--level optimization), Xu and Antsaklis \cite{xu04} prove that the bottom--level, continuous--time optimization is equivalent to a conventional optimal control problem over $I$ fixed intervals with $I$ different system dynamics. In the top--level optimization, the switching--time state is optimized using a gradient descent approach. In order to estimate the gradient of the cost function with respect to the switching times, a set of $I$ Boundary Value Problems (BVPs) are defined. They prove that by iterating over this two alternating optimizations, the switching times and the continuous--time inputs can be optimized effectively. 

In \cite{farshidian16}, we have extended this approach to nonlinear systems with an arbitrary, two times differentiable cost function. In our algorithm, we show that the solution to the top--level optimization BVPs for estimating the gradient can be calculated efficiently using a sweeping method. In this paper, we use the same algorithm to solve the optimal motion control problem for a legged robot system. However, in our case, we need to extend the bottom--level optimization method in order to incorporate the state and input constraints introduced by the contact model and the hybrid system transformation (refer to Equation~\eqref{eq:switched_equations}).    

\subsection{Constrained SLQ Algorithm}
In this section, we first introduce our continuous--time constrained SLQ algorithm. We then use this algorithm to solve the bottom--level optimization with fixed switching times defined in Equation~\eqref{eq:switched_equations}. The proposed optimal control algorithm synthesizes a continuous controller for a nonlinear system subject to a set of state and input constraints. The objective is to minimize an arbitrary, twice--differentiable cost function such as the cost defined in Equation~\eqref{eq:general_op}. To this end, we extend the unconstrained SLQ algorithm to the constrained case. The goal is to devise an algorithm which can handle state and input equality constraints, while having a linear computational complexity with respect to the optimization time horizon as its unconstrained counterpart. This linear--time complexity, in contrast to the cubic--time complexity of the common SQP solver, is an important feature which makes it plausible in an MPC framework on high dimensional systems. 

In \cite{sideris10}, the authors proposed a method for solving a discrete--time constrained SLQ algorithm in $O(n^3)$. Their approach is based on the discrete Bellman equation of optimality. In this work, we extend these results to the continuous--time case while keeping the computational complexity of $O(n)$. Here, $n$ is the average number of points returned by the ODE solver for approximating continuous solutions of differential equations (system dynamics or Riccati equations). It is easy to see that $n$ scales linearly with the optimization time horizon. Furthermore, our proposed approach uses a variable step ODE solver to calculate the optimal controller. As we show in our simulation results, using continuous--time formulation drastically increases the average stepping size of the algorithm while keeping the accuracy of the solution. In practice, this reduces the computational cost of the algorithm especially in its backward pass which is required to compute the costly linearization of the system dynamics. 

The proposed algorithm is an iterative method that in each iteration approximates the nonlinear optimal control problem with a local Linear Quadratic (LQ) subproblem and then solves it through an efficient Riccati based approach \cite{bryson75}. The first step of each iteration is a forward integration of the system dynamics using the last approximation of the optimal controller. Then it calculates a quadratic approximation of the cost function over the state and input nominal trajectories derived from the forward integration. Therefore, for each phase of the switched system optimization we will have
\begin{align}
& \widetilde{J} = \sum_{i=1}^I{ \widetilde{\Phi}_i(\vx({s_i}))+ \int_{s_{i-1}}^{s_i} { \widetilde{L}_i(\vx,\vu,t)dt}} \notag \\
&\widetilde{\Phi}_i(\vx({s_i})) = \ q_{s_i} + \vq_{s_i}^\top \delta\vx + \frac{1}{2} \delta\vx^\top \vQ_{s_i} \delta\vx \notag \\
& \widetilde{L}_i(\vx,\vu,t) = q_i(t) + \vq_i(t)^\top \delta\vx + \vr_i(t)^\top \delta\vu  + \delta\vx^\top \vP_i(t) \delta\vu  \notag \\
& \hspace{17mm} + \frac{1}{2} \delta\vx^\top \vQ_i(t) \delta\vx + \frac{1}{2} \delta\vu^\top \vR_i(t) \delta\vu  \label{eq:cost_quadratic_approximation}
\end{align} 
where $q_i$, $\vq_i$, $\vr_i$, $\vP_i$, $\vQ_i$, and $\vR_i$ are the coefficients of the Taylor expansion of the cost function in Equation~\eqref{eq:general_op} around the nominal trajectories. $\delta\vx$ and $\delta\vu$ are the deviations of state and input from the nominal trajectories. SLQ also uses a linear approximation of the system dynamics and constraints around the nominal trajectories as follows 
\begin{align} 
& \delta\dot{\vx} = \vA_i(t)\delta\vx + \vB_i(t)\delta\vu \notag \\
& \vC_i(t)\delta\vx + \vD_i(t)\delta\vu + \ve_i(t) = \mathbf{0} \notag \\
& \vF_i(t)\delta\vx + \vh_i(t) = \mathbf{0} 
\label{eq:dynamics_linear_approximation}
\end{align}
which are the linear approximations of respectively the system dynamics, state-input constraint, and the pure state constraint in Equation~\eqref{eq:general_op}. Based on this LQ approximation, the SLQ algorithm uses the generalized constrained LQR algorithm introduced in Section~\ref{sec:gen_lqr} to find an update to the feedback-feedforward controller
\begin{equation}
\vu(t,\vx) = \overline{\vu}(t) + \alpha\vl(t) + \alpha_e\vl_e(t) + \vL(t) \left(\vx-\overline{\vx}(t)\right) 
\label{eq:controller}
\end{equation}
Equation~\eqref{eq:controller} gives the update formula for the feedforward--feedback controller. $\overline{\vx}$ and $\overline{\vu}$ are the nominal state and input trajectories. $\vL$, $\vl$, and $\vl_e$ are produced by the constraint SLQ algorithm which are respectively the LQR feedback gains and the feedforward inputs for the cost reduction and the constraint correction. The later will vanish if all the constraints are fulfilled. The parameter $\alpha$ and $\alpha_e$ are the line--search learning rates for the feedforward inputs. These variables are normally determined using a line search scheme. We have implemented two line search methods. The first one assumes that $\alpha=\alpha_e$ and uses a merit function to find the best learning rate. In the other approach, we fix $\alpha_e$ to a small value (e.g. 0.3) and we perform a line search only over $\alpha$ based on the cost function. However, both approaches show a comparable performance. In Section~\ref{sec:gen_lqr}, we will explain the constrained SLQ derivation in more details.

\subsection{Remarks}
Our two-level optimization approach has an interesting connection to the approximate inference methods in machine learning. The approximate inference employs an alternating optimization approach for estimating the latent variables distribution which then is used for maximizing the likelihood function with respect to the model parameter. Similarly in our approach we have defined the switching times as the extended state of the system (similar to a hidden or latent variable) where we estimate them along with the continuous controller parameters in an alternating optimization approach. 

The idea of using a fixed sequence of contacts and adding the switching times to the optimization variables has been proposed by other researchers \cite{koch12,lengagne13}. However, in our approach we are optimizing the switching times through an alternating optimization method. This transforms the continuous control inputs optimization into a conventional optimal control problem which consequently allows us to use an efficient solver such as SLQ. Therefore, this approach can potentially result in a more efficient algorithm.   

Finally, an important aspect of our SLQ algorithm is that the initial solution is not required to be constraint satisfactory. Therefore, the changes of the switching times through the top-level optimizer do not interfere with the performance of the inner loop algorithm even when it causes the initial solution to violate the constraints.   

\section{Generalized Constrained LQR} \label{sec:gen_lqr}
In order to calculate the update rule in the proposed SLQ algorithm, we need to solve a constrained, time--varying LQR problem defined by the cost function in Equation~\eqref{eq:cost_quadratic_approximation} and the system dynamics and constraints introduced in Equation~\eqref{eq:dynamics_linear_approximation}. To solve this constrained optimization problem, we use the method of Lagrange Multipliers \cite{bryson75} and temporally ignore the pure state constraints. Thus, the Lagrangian function for the constrained optimization problem can be written as
\vspace{-1em}

\begin{small}
\begin{equation*}
\begin{split}
    \cL(\vx, \vu, t, \vlambda, \vnu) =  \int_{t_0}^{t_f}  \widetilde{L}(\vx, \vu) + \vlambda^\top (\widetilde{\vf}(\vx, \vu) - \dot \vx) + \vnu^\top \widetilde{\vg_1}(\vx, \vu) \, dt,
\end{split}
\end{equation*}
\end{small}%
where $\vlambda (\cdot)$ and $\vnu (\cdot)$ are Lagrange Multipliers for system dynamics and state--input constraints respectively. The variables denoted with tilde are constructed by consecutively adding the corresponding variables from each mode. We define the Hamiltonian as 
\begin{equation*}
H(\vx, \vu, t, \vlambda, \vnu) = \widetilde{L}(\vx, \vu)  + \vlambda^\top\; \widetilde{\vf}(\vx, \vu) + \vnu^\top \widetilde{\vg_1}(\vx, \vu)
\end{equation*}
and therefore we get
\begin{equation}
\label{eq:LagrHa}
    \cL(\vx, \vu, t, \vlambda, \vnu) = 
    \int_{t_0}^{t_f} 
    H(\vx, \vu, t, \vlambda, \vnu) 
    - \vlambda^\top \dot \vx  \, dt 
\end{equation}
Applying the Euler-Lagrange Equation from the calculus of variations  to Equation~\eqref{eq:LagrHa} leads to
\begin{align}
    \partial_{\vx} H + \dot \vlambda &= 0  \label{eq:dxH} \\
    \partial_{\vu} H &= 0 \label{eq:dvuH} \\
    \partial_{\vnu} H &= 0 \label{eq:dnuH} \\
    \partial_{\lambda} H - \dot \vx&= 0 \label{eq:dlaH}
\end{align}
with the Transversality Condition
\begin{equation}
\vlambda(t_f) = \partial_{\dot \vx} \widetilde{\Phi}(\vx(t_f)) 
\label{eq:lambdaTerm}
\end{equation}
This is essentially the \emph{Pontryagin Minimum Principle} for the constrained linear quadratic optimal control problem that we stated. Inserting the formulation of the stated control problem (Equations~\eqref{eq:cost_quadratic_approximation}~and~\eqref{eq:dynamics_linear_approximation}) into the optimality conditions in Equations~\eqref{eq:dxH} to~\eqref{eq:lambdaTerm} results
\begin{align}
&\vQ \vx + \vP \vu + \vq + \vA^\top \vlambda + \vC^\top \vnu + \dot \vlambda = 0 
\label{eq:cond1}  \\[1ex]
&\vR \vu + \vP^\top \vx + \vr + \vB^\top \vlambda + \vD^\top \vnu = 0  
\label{eq:cond2} \\[1ex]
&\vC \vx +\vD \vu +\ve = 0
\label{eq:cond3} \\[1ex]
&\vA \vx +\vB \vu - \dot \vx = 0
\label{eq:cond5} \\[1ex]
&\vlambda (t_f) - \vQ_f \vx_f -\vq_f = 0
\label{eq:cond6} 
\end{align}
By solving Equations~\eqref{eq:cond2} and \eqref{eq:cond3} together we will have
\begin{align}
    \vnu &= (\vD \vR^{-1} \vD^\top)^{-1}(\vC \vx +\ve ) - 
    [(\vx^\top \vP+ \vlambda^\top \vB+ \vr^\top) \vD^\dagger ]^{\top}
    \notag \\
    \vu &= -(\vI - \vD^\dagger \vD) \vR^{-1}(\vP^\top \vx + \vB^\top \vlambda + \vr )
        -\vD^\dagger(\vC \vx +\ve )
        \label{eq:solve_u_without_nu}
\end{align}
where $\vD^\dagger=\vR^{-1} \vD^\top(\vD \vR^{-1} \vD^\top)^{-1}$ is the right pseudo--inverse of $\vD$. The derived controller in Equation~\eqref{eq:solve_u_without_nu} has an interesting interpretation. Here, the feedforward-feedback controller is composed of two terms. The first term is the unconstrained SLQ controller projected into the null space of the constraint through the null--space projection matrix $(\vI - \vD^\dagger \vD)$ . The second term is a controller which drives the constraint violation to zero. This term is in the range space of the constraint due to the right pseudo--inverse multiplication. After a few reformulations and choosing the following Ansatz $\vlambda (t) = \vS (t)\vx(t) + \vs(t) + \vs_e(t) $. We derive the following equations
\begin{align}
- \dot \vS &= \widetilde\vA^\top \vS + \vS^\top \widetilde\vA - \widetilde\vL^\top \widetilde\vR \; \widetilde\vL + \widetilde\vQ  \hspace{10mm} \vS(t_f)=\vQ_f \label{eq:DEQ_for_Sm}  \\
- \dot \vs &= \widetilde\vA^\top \vs - \widetilde\vL^\top \widetilde\vR \; \widetilde\vl + \widetilde\vq  \hspace{23.5mm} \vs(t_f)=\vq_f  \\
- \dot \vs_e &= \widetilde \vA ^\top \vs_e - \widetilde\vL^\top \widetilde\vR  \; \widetilde\vl_e + ( \widetilde\vC - \widetilde\vL )^\top \vR \; \widetilde\ve \hspace{4mm} \vs_e(t_f)=\boldmath{0}
\end{align}
using the definitions
\begin{align*}
\widetilde\vA &= \vA - \vB \vD^\dagger \vC \\
\widetilde\vC &= \vD^\dagger \vC, \quad \widetilde \vD = \vD^\dagger \vD, \quad \widetilde \ve = \vD^\dagger \ve \\
\widetilde \vQ &= \vQ + \widetilde\vC^\top \vR \; \widetilde\vC  - \vP \widetilde\vC - (\vP \widetilde\vC)^\top \\
\widetilde\vq &= \vq - \widetilde \vC^\top \vr, \quad \widetilde \vR = (\vI - \widetilde \vD)^\top \vR (\vI - \widetilde \vD) \\
\widetilde\vL &= \vR^{-1} ( \vP^\top + \vB^\top \vS ) \\
\widetilde\vl &= \vR^{-1} ( \vr + \vB^\top \vs ), \quad \widetilde \vl_e = \vR^{-1} \vB^\top \vs_e
\end{align*}
The controller feedback and feedforward updates are as
\begin{align}
& \vL = -(\vI - \widetilde\vD) \widetilde\vL - \widetilde\vC  \label{eq:controller_L} \\
& \vl = -(\vI - \widetilde\vD) \widetilde\vl                  \label{eq:controller_l} \\
& \vl_e =  -(\vI - \widetilde\vD) \widetilde\vl_e - \widetilde\ve  \label{eq:controller_le} \\
& \vu =  \vl + \vl_e + \vL \vx  \label{eq:lqr_controller}
\end{align}
Equation~\eqref{eq:lqr_controller} gives the update formula for the controller. 

We will now consider the case of pure state constraint. As shown by \cite{sideris10}, using a Lagrange multiplier approach to handle this type of constraint results in an $O(n^3)$ algorithm. In order to avoid this, we augment the cost function with a term to penalize the constraints violation. To do so, we choose the integral of squared norm error criterion. However, a naive implementation of this method needs the second order derivatives of the constraints. Since these constraints are normally a function of the robot kinematics and computing the second order derivatives is in general expensive, we choose a method similar to the nonlinear least square approach where we use the normal equation of the pure state constraint instead of the second order approximation. The normal equation for the pure state constraints is defined as 
\begin{equation} \label{eq:normal_equation}
\vF^\top \vF \vx + \vF^\top \vh = 0
\end{equation} 
We add this normal equation through a penalty coefficient to the state derivative of the Hamiltonian in Equation~\eqref{eq:dxH} which corresponds to the variational of the Lagrangian with respect to state. To do so, we should modify $\widetilde\vQ$ and $\widetilde\vq$ by adding an extra term $\vF^\top \vF$ and $\vF^\top \vh$ respectively. By eventually increasing the penalty multiplier over iterations, we can enforce the pure state constraints to desired accuracy. 

This concludes the necessary conditions of the optimality. For the sufficient conditions, we should show that there exists a function which satisfies the Hamilton--Jacobi--Bellman equation. It is easy to show that $V(t,\vx)=\frac{1}{2}\vx^\top \vS(t) \vx + \vx^\top \vs(t) + s(t)$ satisfies this equation where $s(t)$ is defined as
\begin{equation} \label{eq:DEQ_for_s}
-\dot{s} = q - \widetilde\vl^\top \widetilde\vR \; \widetilde\vl, \qquad s(t_f)=q_f
\end{equation} 
%

\section{Motion Control for the CoM and the Full Kinematics}
In order to validate our algorithm on the legged robot application, we use a switched model formulation in Equation~\eqref{eq:switched_equations} for modeling the CoM dynamics plus full kinematics \cite{dai14}. The quadruped robot used for the simulation is our hydraulicly actuated robot, HyQ, with 3 degrees of freedom per leg \cite{hyq11}. The generalized torque in Equation~\eqref{eq:switched_equations} is set to zero since the only effective forces on the CoM are the external forces. The CoM inertia tensor can be calculated using the inertia tensor and the CoM position of each link. The CoM plus kinematics model has in total $24$ states consisting of: $3$ states for base orientation, $\vth$, $3$ states for CoM position, $\vp$, $6$ states for linear and angular velocities of CoM in body frame, $\vv$  and $\vom$, and $12$ joint angels $\vq$. The control inputs of the model are switching times, contact forces of legs $\{\vlambda^i(t)\}_{i=1}^4$, and joint angular velocities $\vu$. Using the Newton-Euler equation for modeling the CoM dynamics, we get   
\vspace{-1em}

\begin{small}
\begin{align}
&\left\{ 
\begin{array}{ll}
		\dot{\vth} = \vR \left( \vom - _B\vJ_{CoM}^{\omega} \, \dot\vq \right) \\
		\dot{\vp}  = \vR \, \vv \\
		\dot{\vom} = \vI^{-1} \left( \dot\vI -\vom \times \vI\vom + \sum_{i=0}^4{ \vr_{f_i} \times \vlambda_i} \right)\\
		\dot{\vv}  = \vg + \frac{1}{m} \sum_{i=0}^4{\vlambda_i} \\
		\dot{\vq} = \vu 
\end{array}	 
\right. \notag \\
&\left\{ 
\begin{array}{ll}
		\vu_{f_i} = \mathbf{0}      \hspace{15mm}  \text{for stance legs} \\
		\vlambda_{f_i} = \mathbf{0}  \hspace{15mm} \text{for swing legs} \\
		\vu_{f_i} \cdot \hat{n} = c(t)   \hspace{8mm}    \text{for swing legs} \\
\end{array}	 
\right.
\end{align}
\end{small}%
where $\vR$ is the rotation matrix of the base with respect the global frame, $\vg$ is the gravitational acceleration in body frame, $\vI$ and $m$ are moment of inertia about the CoM and the total mass respectively. $_B\vJ_{CoM}^{\omega}$ is the Jacobian matrix of CoM rotation with respect to robot base frame. $\vr_{f_i}$, $\vu_{f_i}$, $\vlambda_{f_i}$ are respectively the position, velocity and contact force vector of foot $i$. We use a predefined swing leg trajectory ($c(t)$) in the orthogonal direction of the contact surface ($\hat{n}$) which ensures that the touch-down takes place according to the switching time schedule and the given mode switch sequence. Furthermore, it ensures that the velocity of the swing foot before the contact is zero. The unilateral constraints on contact forces are enforced by considering an extra equality constraint $\vlambda_f=\mathbf{0}$ for any leg that violates the constraint during the forward pass of SLQ.  

One interesting aspect of our modeling is the use of the base orientation instead of the CoM orientation. This choice has a practical motive rooting from the fact that the CoM orientation is not a measurable physical entity. Therefore, using its value in a feedback controller structure requires an estimation scheme through integrating CoM velocity which is prone to drift in the absence of direct measurements.   

\section{Results}
In order to validate our approach, we present three tasks using different contact patterns, i.e. different gaits. First, we demonstrate a walking gait in a free environment. Afterwards, a ``gap'' is introduced which constrains footholds to lie outside of it. As a final test, we demonstrate trotting in different directions. During all tests, contact switching timings are initialized uniformly in time. The top--level optimizer then finds optimized switching times according to the task at hand. The reader can find the videos of the following simulation results online\footnote{\url{https://youtu.be/KHi_C-SsC2A}}.

\subsection{Walking Task}
\begin{figure}[tbp]
\centering
\includegraphics[width=\columnwidth]{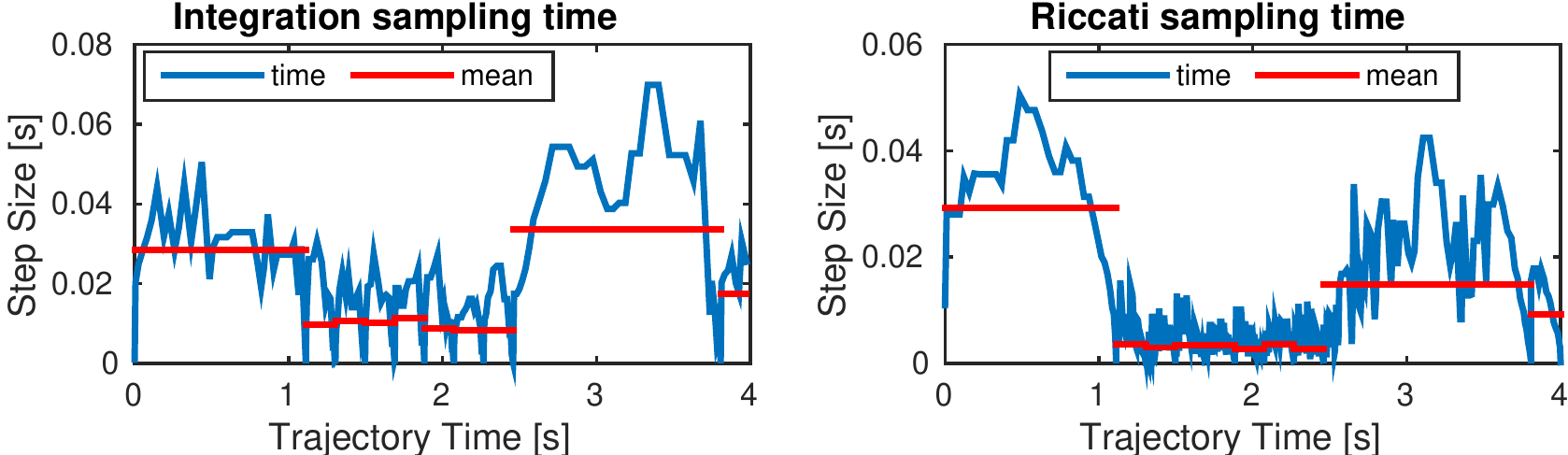}
\caption{Plot of the step sizes for integration and solving the Riccati equation in the walking task. Both step sizes are higher in the initial four legged support phase and the second last phase.}
\label{fig:walking_sampling}
\end{figure}

In this task, we specify a lateral sequence walking which three feet are always on the ground at the same time. Fig.~\ref{fig:walking_sampling} shows the corresponding SLQ step--sizes for each phase. The average step sizes in the forward and backward passes are significantly higher in comparison to what we normally choose in the discrete--time SLQ (around 0.002). This is especially prominent in the initial four legged support phase and the second last phase. Since these are also the longest phases, significant computational effort can be saved compared to a fixed step--size implementations such as the discrete--time SLQ algorithm.
\begin{figure*}[tbp]
\centering
\includegraphics[width=2\columnwidth]{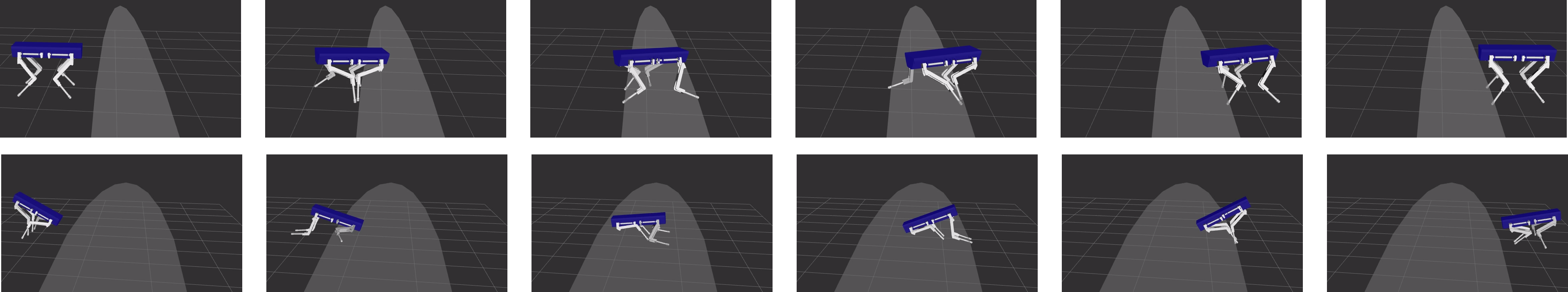}
\caption{Time series of the 1~m gap crossing using a walking gait and of a jump over a 2~m gap. When crossing a 1~m gap, HyQ almost reaches its kinematic limits, avoiding it by slightly rotating its body allowing it to use both its hip degrees of freedom simultaneously. In the case of the 2~m gap, HyQ has to execute a rather extreme jump, leaping over the gap.}
\label{fig:time_series}
\end{figure*}

\subsection{Gap Crossing Task} 
In the next task, we apply the same walking gait to a ``gap'' crossing problem. In this scenario, HyQ has to reach a target point beyond a wide gap. This gap defines an area in which HyQ is not allowed to place its feet. This gap is \\
\begin{figure}[H]
\centering
\includegraphics[width=\columnwidth]{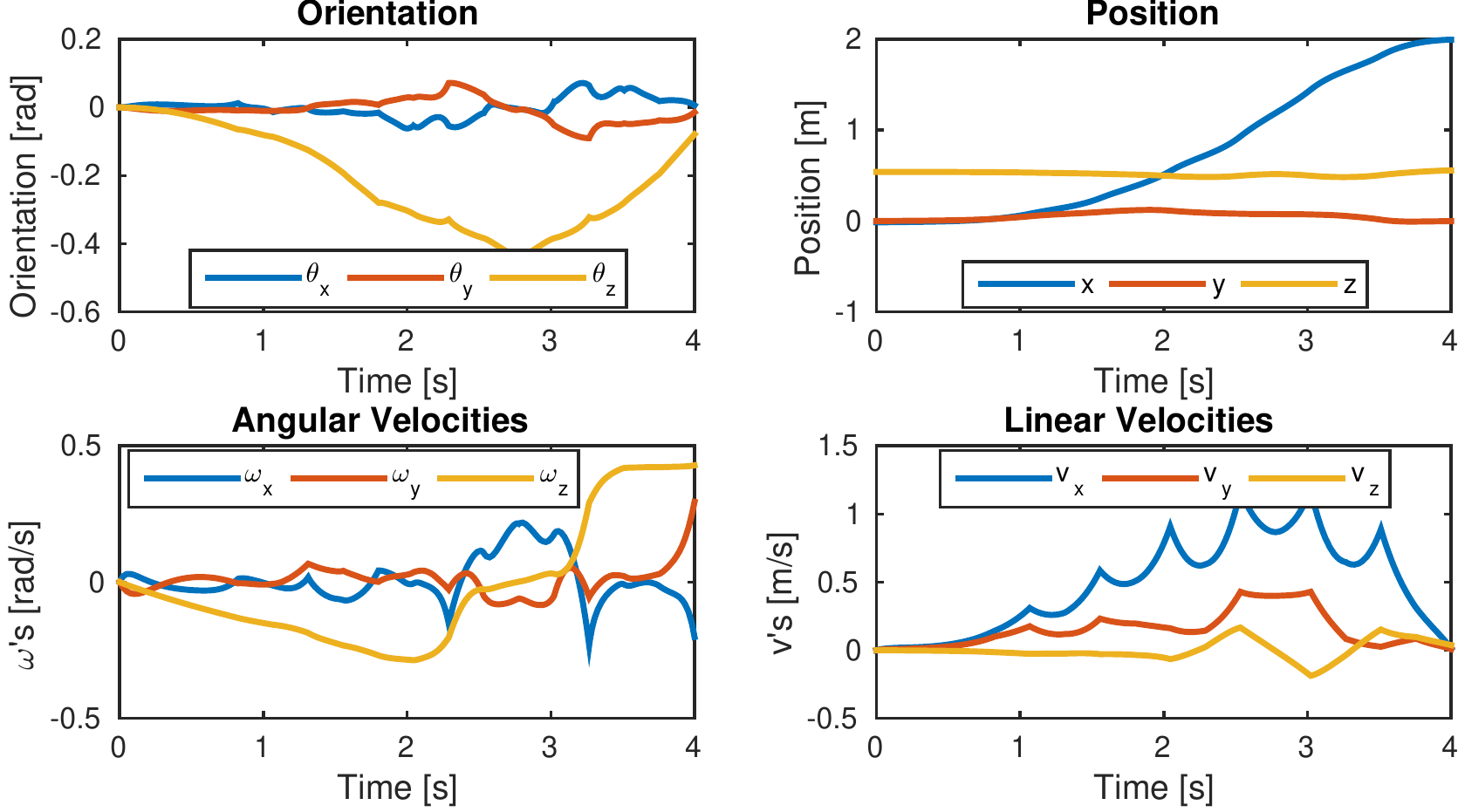}
\caption{Center of Mass pose plots for HyQ crossing a 1m gap. Interestingly, the trajectory contains a yaw motion which effectively increases the kinematic reach of the leg.} 
\label{fig:gap_crossing_1m_base}
\end{figure}
\vspace*{-7mm}
\begin{figure}[H]
\centering
\includegraphics[width=\columnwidth]{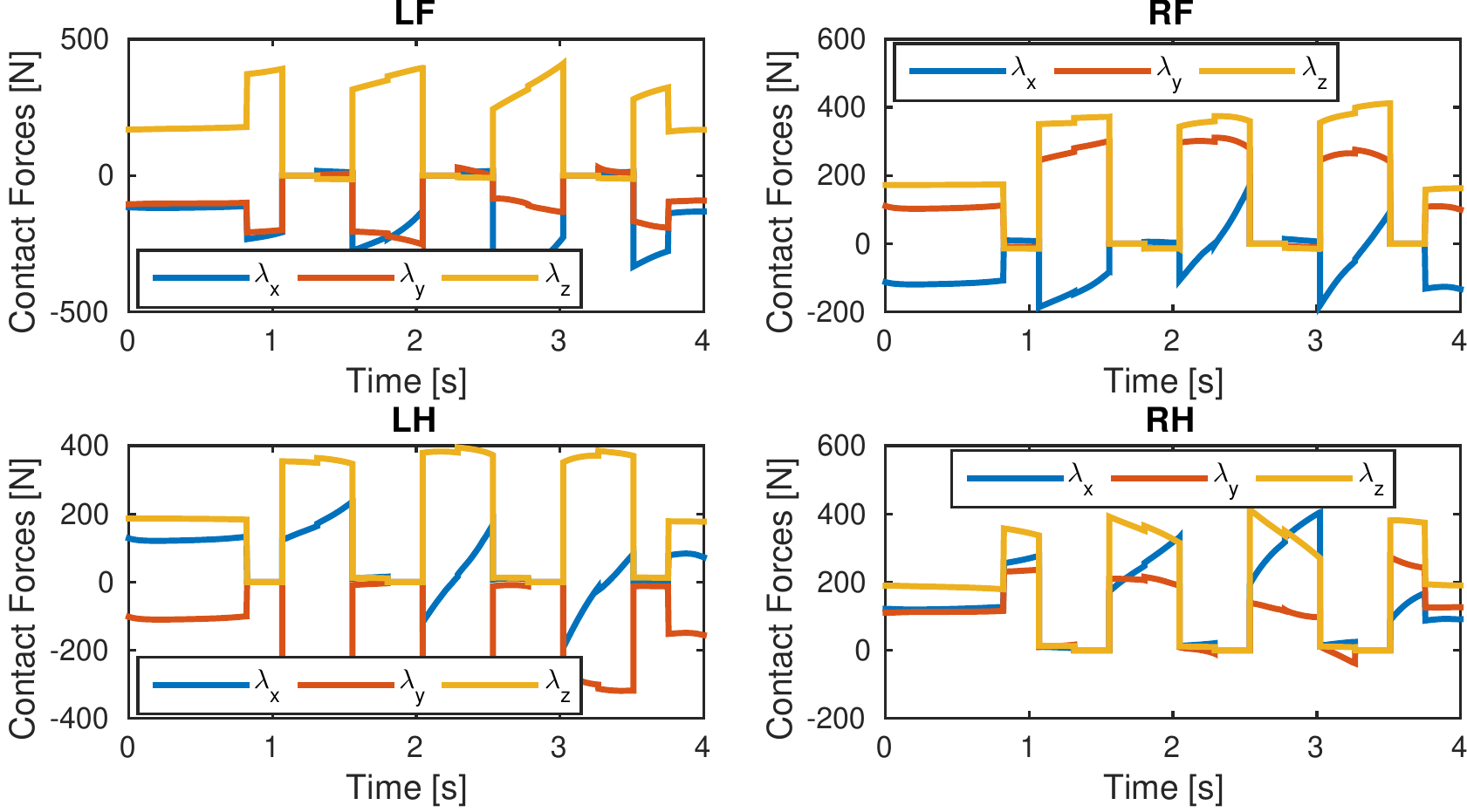}
\caption{Contact forces for the 1m gap crossing task. The contact forces are relatively smooth during support phases, which will potentially facilitate hardware experiments. The contact forces also nicely reflect the stepping pattern during the task.}
\label{fig:gap_crossing_1m_forces}
\end{figure}
\vspace*{-4mm} \noindent
encoded as a potential field and included as a state constraint. This constraint is only enforced at the end of each swing leg phase, i.e. at touchdown of the leg using the terminal cost of modes. As we can see from the motion sequence illustrated in  Fig.~\ref{fig:time_series}, the optimization finds a gait which avoids stepping onto the gap. Since the gap is relatively wide compared to the robot's kinematic limits, HyQ yaws and crosses the gap in a slight diagonal configuration, visible in Fig.~\ref{fig:gap_crossing_1m_base}. This allows to use the hip degree of freedom for increased mobility. This behavior is probably not expected at first, but underlines the potential of our approach. 

\begin{figure}[tbp]
\centering
\includegraphics[width=\columnwidth]{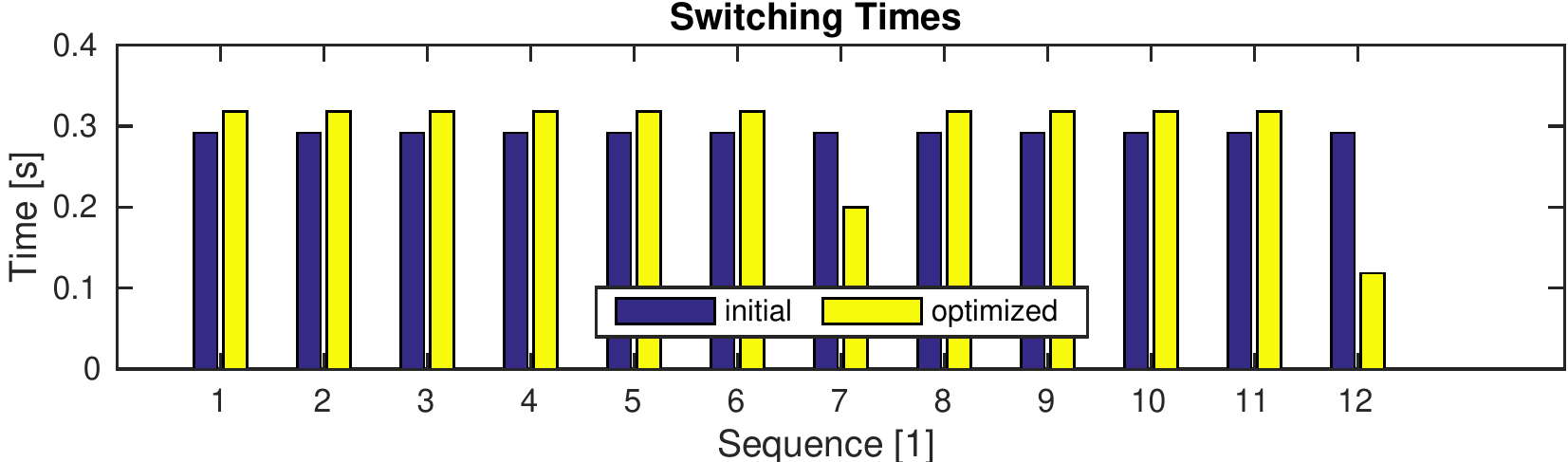}
\caption{Switching times for jumping over a 2~m gap. Sequence 7 corresponds to the flight sequence. To avoid unnecessary high jumps, the top-level optimizer reduces the time of the sequence to a minimum to clear the gap.}
\label{fig:gap_jumping_2m_switching_times}
\end{figure}

\vspace*{1mm}
We alter the gap crossing task by increasing the gap size to 2~m. We also use a trotting gait instead of walking in order to gain more speed to jump over the gap. Furthermore, we add a flight phase with no contacts in between, since the gap is wider than the reach of the HyQ's leg. The optimization algorithm automatically places the flight phase over the gap, such that HyQ executes a jump, as shown in Fig.~\ref{fig:time_series}. Fig.~\ref{fig:gap_jumping_2m_switching_times} compares initial and optimized switching times for the task. The top-level optimization shortens the flight phase sequence. This effectively reduces flight time and thus the jump height to a lower optimum. 

\vspace*{1mm}
Fig.~\ref{fig:gap_crossing_1m_forces} shows the contact forces during the gap crossing motion. The contact forces nicely visualize the stepping sequence where the discontinuities in the contact forces concur with  a moment of touch--down or lift--off. Furthermore, The contact force are relatively smooth during support phases and there are no distinct spikes during touch--down or lift--off. Thus, when implementing the motions on hardware, they provide smooth references for tracking controllers which will potentially help with robustness during experiments.

\subsection{Forward and Side Trotting Task}
\begin{figure}[tbp]
\centering
\includegraphics[width=\columnwidth]{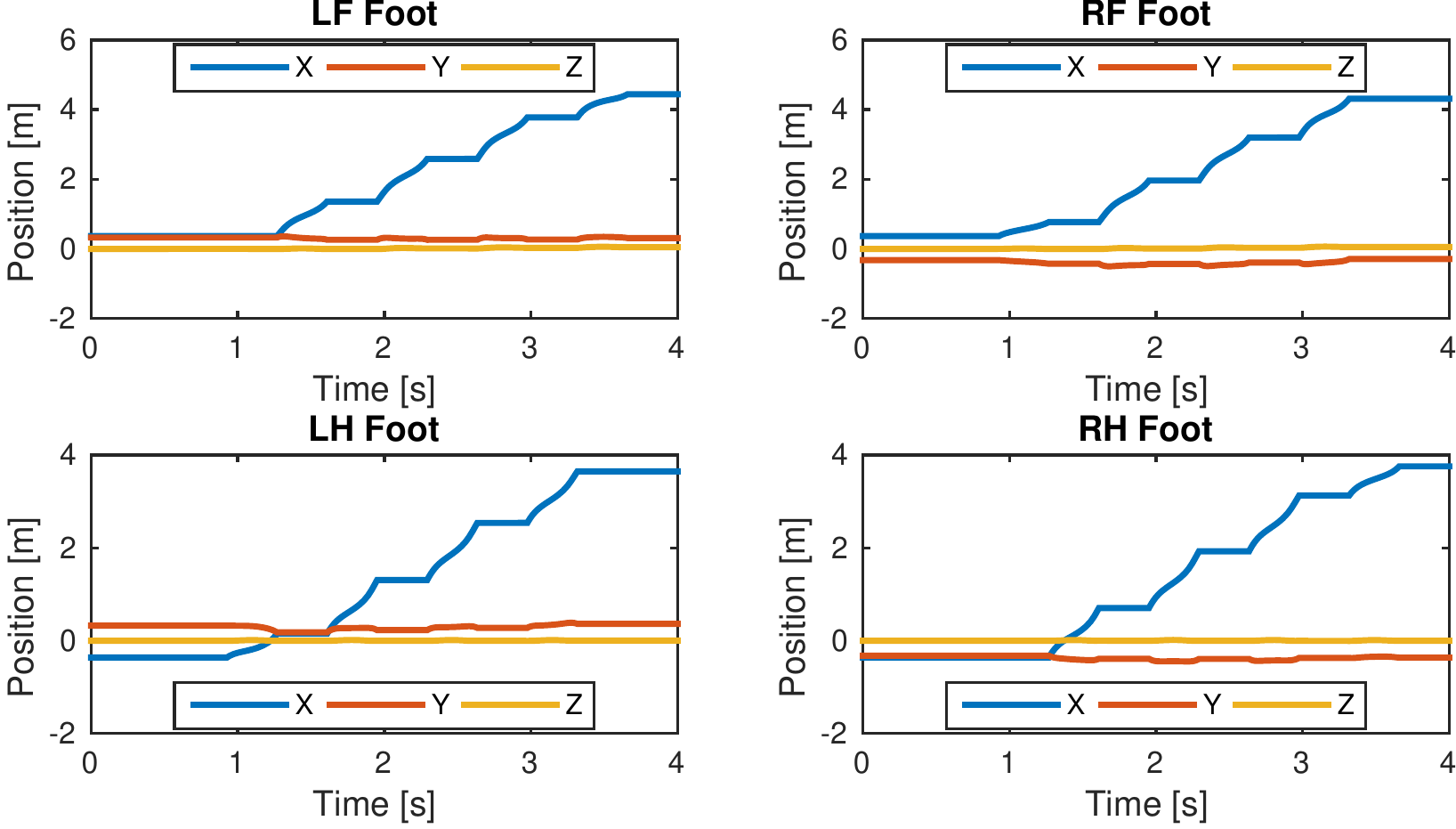}
\caption{Plot of feet positions for the forward trotting task. The step length in x-direction is even throughout the trajectory, with the exception of acceleration and deceleration phases. There is no significant side stepping and no ground penetration, underlining good constraint satisfaction.}
\label{fig:trotting_feet_pos}
\end{figure}

In order to demonstrate that our approach also translates to more dynamic, statically unstable modes, we apply it to a trotting gait. Here, always one diagonal leg pairs is in contact with the ground, without a four leg support phase in--between. By altering the cost function, we can set different target points for the robot to reach. This way we can obtain a forward and a lateral trot. In Fig.~\ref{fig:trotting_feet_pos} we can see the feet positions for the forward trot. We see that the step length is even throughout the trajectory. Furthermore, there is no pronounced side stepping. Lastly, we also see that there is no ground penetration, which underlines the low constraint violation also shown in Fig.~\ref{fig:cost_ise}.

\subsection{Convergence and Constraint Violation}
So far, we have seen the capabilities of the approach in optimizing motions for each tasks. In this test, we analyze how quickly these motions are found and if they violate constraints. Additionally, we observe how much can be gained from the switching time optimization.
\begin{figure}[tbp]
\centering
\includegraphics[width=\columnwidth]{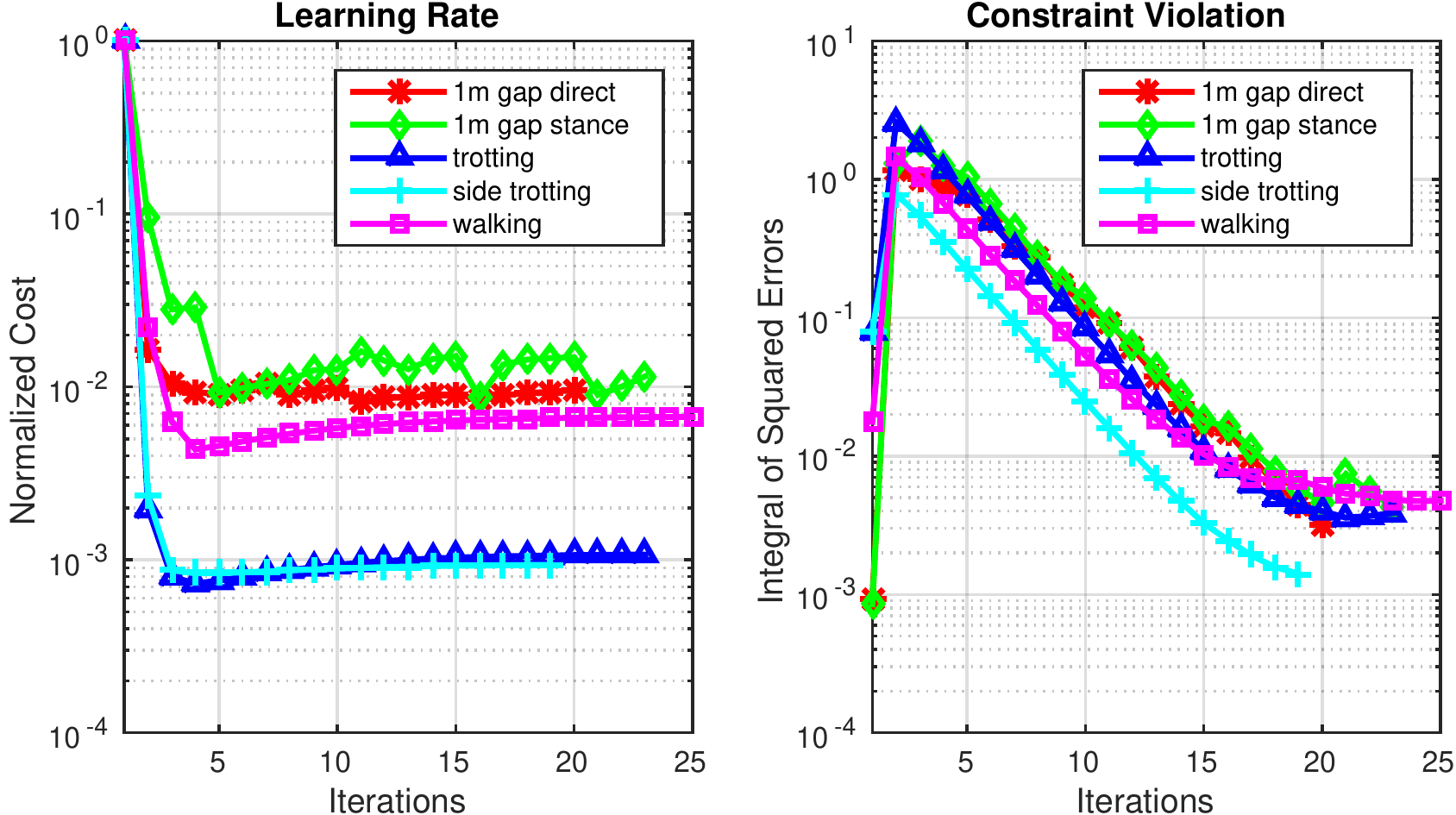}
\caption{Plot of normalized costs over iterations as well as constraint violation expressed as the Integral of Squared Errors over iterations. The optimizer first reduces the costs in the first few iterations. Afterwards it fixes the constraint violation while the costs remain at a constant low level.}
\label{fig:cost_ise}
\end{figure}
Fig.~\ref{fig:cost_ise} shows the decrease of costs over iterations of the inner optimization, as well as the constraint violation expressed as the Integral of Squared Errors (ISE) over iterations. The first iteration is always initialized with a simple standing controller, leading to bad costs but relatively good constraint satisfaction. After one or two iterations, costs are significantly reduced. However, this cost reduction comes at the expense of increased constraint violation. Over the course of the following iterations, the constraint violation is reduced while costs remain almost constant. After about 10 to 15 iterations, the ISE of the constraint violation is reduced to a very low range. In this range, physical inaccuracies and limitations of a physical system probably play a more important role during execution than the constraint violation. Furthermore, we can see that constraint violation decreases exponentially (linear on logarithmic scale) over the iterations, independent of the given task. This underlines the robust convergence behavior of the approach.
\begin{figure}[tbp]
\centering
\includegraphics[width=\columnwidth]{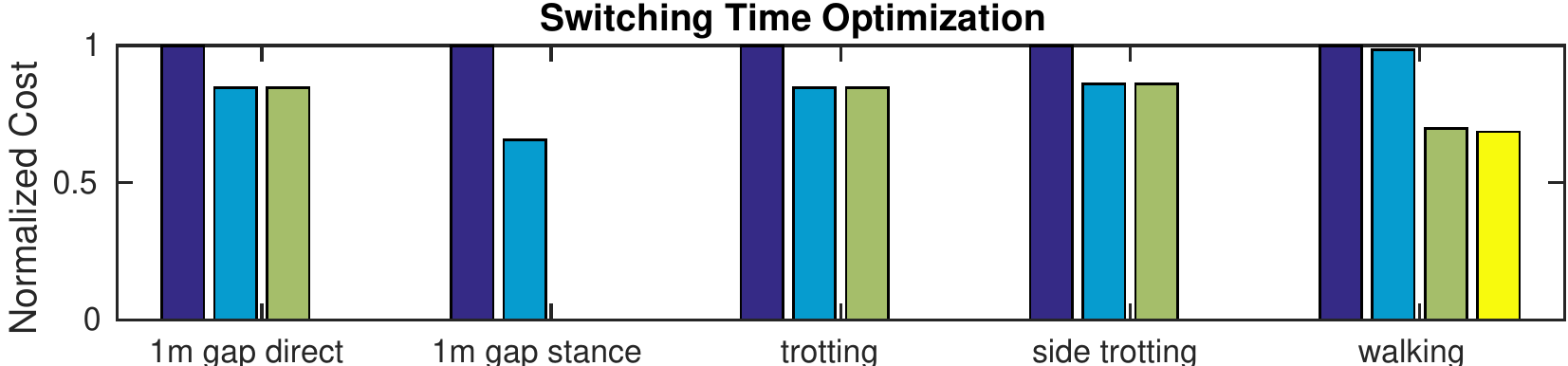}
\caption{Normalized costs for different tasks during different iterations of the top-level optimization. This optimization is both effective and efficient. It reduces the costs by up to 40\% in usually 2-3 iterations. Each bar shows one iteration.}
\label{fig:outer_iterations}
\end{figure}

Fig.~\ref{fig:outer_iterations} shows the normalized costs over the top-level optimization iterations for different tasks. There are two interesting observations to be made: First, by optimizing the switching times, costs can be reduced by 15 to 40\%. Secondly, this reduction is usually achieved in 2-3 iterations. This underlines that the outer switching time optimization is very effective but also efficient. The results in this section demonstrate that both levels of the algorithm converge rapidly in a wide range of tasks. This supports our previous discussion that the two--level optimization problem decomposes the original problem into two optimizations which can be solved more efficiently.

\section{Conclusion and Outlook}
In this work, we investigated a framework for legged robot optimal motion planning and control. The main contributions of our approach are: switch system modeling for the legged robot, a multi--level optimization approach for optimizing the switching time and the continuous control inputs, and finally the development of a constrained SLQ algorithm. Our switched system formulation facilitates us to use more efficient tools for solving the optimal control problem in legged robots. In this paper, we have chosen a multi--level optimization approach to find the switching times and the continuous control inputs. As we have shown in our experiments, this optimization approach converges rapidly on both levels. This manifests the discussion that the two--level optimization framework decomposes the problem in two simpler optimizations which potentially can be solved more efficiently with existing algorithms. 

Finally in this work, for the first time we proposed a continuous--time, constrained DP approach based on the SLQ framework. This algorithm similar to its unconstrained ancestors has $O(n)$ time--complexity with the advantage that it can also handle state and input equality constraints. This work not only extends the state-of-the-art available SLQ algorithm to the constrained problems but also introduces a new constrained, time--varying LQR solver. This constrained LQR method can be also used for designing the feedback controller for the open--loop planners such as the ones in the TO framework.    

One of the immediate extensions to our constrained SLQ algorithm is to use an active set approach in order to incorporate the state and input inequality constraints. Furthermore, we are planning to implement this approach on our robot, HyQ, using an MPC framework.


\section*{Acknowledgment} \footnotesize{This research has been supported in part by a Max-Planck ETH Center for Learning Systems Ph.D. fellowship to Farbod Farshidian and a Swiss National Science Foundation Professorship Award to Jonas Buchli and the NCCR Robotics.}

\bibliographystyle{bibliography/IEEEtran} \bibliography{bibliography/references}

\end{document}

%% file: mathdef.tex
\newcommand{\argmin}{\mathop{\mathrm{argmin}}}

\newcommand{\vth}{\mbox{\boldmath $\theta$}}

\newcommand{\vlambda}{\mbox{\boldmath $\lambda$}}

\newcommand{\vnu}{\mbox{\boldmath $\nu$}}

\newcommand{\vtau}{\mbox{\boldmath $\tau$}}

\newcommand{\vom}{\mbox{\boldmath $\omega$}}

\newcommand{\vd}{\mathbf d}
\newcommand{\ve}{\mathbf e}
\newcommand{\vf}{\mathbf f}
\newcommand{\vg}{\mathbf g}
\newcommand{\vh}{\mathbf h}

\newcommand{\vl}{\mathbf l}

\newcommand{\vp}{\mathbf p}
\newcommand{\vq}{\mathbf q}
\newcommand{\vr}{\mathbf r}
\newcommand{\vs}{\mathbf s}

\newcommand{\vu}{\mathbf u}
\newcommand{\vv}{\mathbf v}

\newcommand{\vx}{\mathbf x}

\newcommand{\vA}{\mathbf A}
\newcommand{\vB}{\mathbf B}
\newcommand{\vC}{\mathbf C}
\newcommand{\vD}{\mathbf D}

\newcommand{\vF}{\mathbf F}

\newcommand{\vI}{\mathbf I}
\newcommand{\vJ}{\mathbf J}

\newcommand{\vL}{\mathbf L}
\newcommand{\vM}{\mathbf M}

\newcommand{\vP}{\mathbf P}
\newcommand{\vQ}{\mathbf Q}
\newcommand{\vR}{\mathbf R}
\newcommand{\vS}{\mathbf S}

\newcommand{\cL}{\mathcal{L}}




%% file: p_17_icra_switched_hyq.bbl
\begin{thebibliography}{10}
\providecommand{\url}[1]{#1}
\csname url@rmstyle\endcsname
\providecommand{\newblock}{\relax}
\providecommand{\bibinfo}[2]{#2}
\providecommand\BIBentrySTDinterwordspacing{\spaceskip=0pt\relax}
\providecommand\BIBentryALTinterwordstretchfactor{4}
\providecommand\BIBentryALTinterwordspacing{\spaceskip=\fontdimen2\font plus
\BIBentryALTinterwordstretchfactor\fontdimen3\font minus
  \fontdimen4\font\relax}
\providecommand\BIBforeignlanguage[2]{{%
\expandafter\ifx\csname l@#1\endcsname\relax
\typeout{** WARNING: IEEEtran.bst: No hyphenation pattern has been}%
\typeout{** loaded for the language `#1'. Using the pattern for}%
\typeout{** the default language instead.}%
\else
\language=\csname l@#1\endcsname
\fi
#2}}

\bibitem{mayne66}
D.~Mayne, ``A second-order gradient method for determining optimal trajectories
  of non-linear discrete-time systems,'' \emph{International Journal of
  Control}, 1966.

\bibitem{todorov05}
E.~Todorov and W.~Li, ``A generalized iterative lqg method for locally-optimal
  feedback control of constrained nonlinear stochastic systems,'' in
  \emph{Proceedings of the 2005, American Control Conference.}, 2005.

\bibitem{sideris05}
A.~Sideris and J.~E. Bobrow, ``An efficient sequential linear quadratic
  algorithm for solving nonlinear optimal control problems,'' in \emph{American
  Control Conference, 2005. Proceedings of the 2005}, 2005.

\bibitem{von92}
O.~Von~Stryk and R.~Bulirsch, ``Direct and indirect methods for trajectory
  optimization,'' \emph{Annals of operations research}, 1992.

\bibitem{diehl06}
M.~Diehl, H.~G. Bock, H.~Diedam, and P.-B. Wieber, ``Fast direct multiple
  shooting algorithms for optimal robot control,'' in \emph{Fast motions in
  biomechanics and robotics}.\hskip 1em plus 0.5em minus 0.4em\relax Springer,
  2006.

\bibitem{enright92}
P.~J. Enright and B.~A. Conway, ``Discrete approximations to optimal
  trajectories using direct transcription and nonlinear programming,''
  \emph{Journal of Guidance, Control, and Dynamics}, 1992.

\bibitem{koch12}
K.~H. Koch, K.~Mombaur, and P.~Soueres, ``Optimization-based walking generation
  for humanoid robot,'' \emph{IFAC Proceedings}, 2012.

\bibitem{lengagne13}
S.~Lengagne, J.~Vaillant, E.~Yoshida, and A.~Kheddar, ``Generation of
  whole-body optimal dynamic multi-contact motions,'' \emph{The International
  Journal of Robotics Research}, 2013.

\bibitem{posa14}
M.~Posa, C.~Cantu, and R.~Tedrake, ``A direct method for trajectory
  optimization of rigid bodies through contact,'' \emph{The International
  Journal of Robotics Research}, 2014.

\bibitem{pardo15}
D.~Pardo, M.~Neunert, A.~W. Winkler, and J.~Buchli, ``Projection based whole
  body motion planning for legged robots,'' \emph{arXiv preprint
  arXiv:1510.01625}, 2015.

\bibitem{ihmc13}
T.~Koolen, J.~Smith, G.~Thomas, S.~Bertrand, J.~Carff, N.~Mertins, D.~Stephen,
  P.~Abeles, J.~Englsberger, S.~Mccrory, \emph{et~al.}, ``Summary of team
  ihmc's virtual robotics challenge entry,'' in \emph{13th IEEE-RAS
  International Conference on Humanoid Robots}.\hskip 1em plus 0.5em minus
  0.4em\relax IEEE, 2013.

\bibitem{winkler17}
A.~W. Winkler, F.~Farshidian, M.~Neunert, D.~Pardo, and J.~Buchli, ``Online
  walking motion and contact optimization for quadruped robots,'' in \emph{IEEE
  International Conference on Robotics and Automation (ICRA)}, 2017.

\bibitem{tassa12}
Y.~Tassa, T.~Erez, and E.~Todorov, ``Synthesis and stabilization of complex
  behaviors through online trajectory optimization,'' in \emph{IEEE/RSJ
  International Conference on Intelligent Robots and Systems}, 2012.

\bibitem{neunert16}
M.~Neunert, F.~Farshidian, A.~W. Winkler, and J.~Buchli, ``Trajectory
  optimization through contacts and automatic gait discovery for quadrupeds,''
  \emph{arXiv preprint arXiv:1607.04537}, 2016.

\bibitem{ohno78}
K.~Ohno, ``A new approach to differential dynamic programming for discrete time
  systems,'' \emph{IEEE Transactions on Automatic Control}, vol.~23, no.~1, pp.
  37--47, 1978.

\bibitem{yakowitz86}
S.~Yakowitz, ``The stagewise kuhn-tucker condition and differential dynamic
  programming,'' \emph{IEEE transactions on automatic control}, vol.~31, no.~1,
  pp. 25--30, 1986.

\bibitem{romano15}
F.~Romano, A.~Del~Prete, N.~Mansard, and F.~Nori, ``Prioritized optimal
  control: A hierarchical differential dynamic programming approach,'' in
  \emph{Robotics and Automation (ICRA), 2015 IEEE International Conference
  on}.\hskip 1em plus 0.5em minus 0.4em\relax IEEE, 2015, pp. 3590--3595.

\bibitem{farshidian16}
F.~Farshidian, M.~Kamgarpour, D.~Pardo, and J.~Buchli, ``Sequential linear
  quadratic optimal control for nonlinear switched systems,'' \emph{arXiv
  preprint arXiv:1609.02198}, 2016.

\bibitem{sideris10}
A.~Sideris and L.~A. Rodriguez, ``A riccati approach to equality constrained
  linear quadratic optimal control,'' in \emph{Proc. American Control
  Conference}, 2010.

\bibitem{xu04}
X.~Xu and P.~J. Antsaklis, ``Optimal control of switched systems based on
  parameterization of the switching instants,'' \emph{IEEE transactions on
  automatic control}, 2004.

\bibitem{bryson75}
A.~E. Bryson, \emph{Applied optimal control: optimization, estimation and
  control}.\hskip 1em plus 0.5em minus 0.4em\relax CRC Press, 1975.

\bibitem{dai14}
H.~Dai, A.~Valenzuela, and R.~Tedrake, ``Whole-body motion planning with
  centroidal dynamics and full kinematics,'' in \emph{IEEE-RAS International
  Conference on Humanoid Robots}, 2014.

\bibitem{hyq11}
C.~Semini, N.~G. Tsagarakis, E.~Guglielmino, M.~Focchi, F.~Cannella, and D.~G.
  Caldwell, ``Design of hyq--a hydraulically and electrically actuated
  quadruped robot,'' \emph{Proceedings of the Institution of Mechanical
  Engineers, Journal of Systems and Control Engineering}, 2011.

\end{thebibliography}
